\documentclass[reprint,showpacs,showkeys,amsmath,amssymb,aps,nofootinbib,prd]{revtex4-1}
\usepackage{graphicx}
\usepackage{epstopdf}
\begin{document}
\title{
Cosmological particle creation in a hadronic fluid
 }
 \author{Neven Bili\'c}
\email{bilic@irb.hr}
\author{Dijana Toli\'c}
\email{dijana.tolic@irb.hr}
\affiliation{
Division of Theoretical Physics, \\ Rudjer Bo\v{s}kovi\'{c} Institute,
P.O.\ Box 180, 10001 Zagreb, Croatia
}
\date{\today}

\begin{abstract}

Acoustic perturbations in an expanding hadronic fluid 
at temperatures below the chiral transition point represent massless pions
propagating  in curved spacetime geometry.
In comoving coordinates  the corresponding analog metric tensor
describes a hyperbolic Friedmann-Robertson-Walker (FRW) spacetime. 
We study the analog cosmological particle creation of pions  
below the critical point of the chiral phase transition.
We compare the cosmological creation spectrum with the
spectrum of analog Hawking radiation
at the analog trapping horizon.
\end{abstract}


\maketitle

\section{Introduction}

\label{introduction}

Quantum field theory in a curved spacetime 
predicts  that the gravitation field 
creates particles and antiparticles.
The main related phenomena are the Hawking effect in black holes 
and particle creation due to the cosmological expansion \cite{parker1}.
In the latter case the creation is caused by the time dependence
of the background metric which in turn
causes a nontrivial time evolution of the ground state similar to
what happens to a quantum harmonic oscillator with time-dependent frequency
\cite{jacobson1}.
The process of particle creation depends only
on the particulars of the metric and particle properties (mass, spin etc.) 
irrespective of whether the expansion is of cosmological or
another origin. In particular, particle creation is expected
in time-dependent analog gravity systems such as expanding
Bose-Einstein (BE) condensates \cite{barcelo1,jain,kurita,sabin},
Bose and Fermi superfluids \cite{fedichev}, and
expanding hadronic fluids 
\cite{tolic,tolic2}.

Analog gravity has proven to be useful   in studying 
various physical phenomena \cite{barcelo}, e.g., 
acoustics \cite{visser},
optics \cite{philbin},
superfluidity
\cite{jacobson2}, black hole accretion \cite{moncrief,abraham},
and hadron fluid
\cite{tolic,tolic2,tolic3}.
The purpose of this paper is to study in the framework 
of analog gravity the phenomenon of cosmological particle creation
in a hadronic fluid produced in high energy collision experiments.

Strongly interacting matter is described at the fundamental level by
a non-Abelian gauge theory called
 quantum chromodynamics (QCD). 
 At low energies, the QCD vacuum is characterized by 
a nonvanishing expectation value
\cite{shifman}:
  $\langle \bar\psi\psi\rangle \approx$ (235 
MeV)$^3$,
theso-called chiral condensate.
This quantity
describes the density of quark-antiquark pairs
found in the QCD vacuum and its nonvanishing value 
is a manifestation of chiral symmetry breaking
\cite{harris}.
Our approach is based on
 the linear sigma model \cite{gell} combined with
 a  boost invariant Bjorken-type spherical expansion \cite{lampert}.
The linear sigma model  serves as an effective
model for the low-temperature
phase of QCD \cite{bilic,bilic1}.
In the chirally broken phase, i.e.,  
at temperatures below
the point of the chiral phase transition,
the pions are massless but owing to the 
finite temperature effects
propagate slower than light
\cite{pisarski,son1,son2}. Moreover, the pion velocity approaches zero at the
critical temperature.

As in general relativity, a notable manifestation of analog gravity are two effects both having 
a quantum origin:
the Hawking radiation and cosmological particle creation.
These two phenomena are similar but appear under different physical
conditions. 
The Hawking thermal radiation is due to the information loss across the apparent horizon
whereas the cosmological particle creation generates a quasithermal radiation 
as a result of time variation of the spacetime geometry. 
The Hawking radiation takes place only if there exists
a trapping (or apparent) horizon,
whereas the cosmological particle creation
takes place with or without a horizon.
On the other hand, in any stationary geometry the cosmological particle creation is absent 
whereas the Hawking radiation is present in a stationary geometry with an event horizon. 
Besides, if a  field theory is conformally invariant there will be no 
cosmological particle creation.  In contrast, the Hawking  radiation is present  
even in the conformal case as long as a trapping horizon exists.
 
The analog Hawking effect has been studied in our previous papers \cite{tolic,tolic2} 
in the context of an expanding hadronic fluid. We have demonstrated 
that there exists a region where the flow velocity exceeds the pion velocity
and the analog trapped region forms which then
causes the Hawking
radiation of massless pions.
Here we study the effect of cosmological creation of pions 
in an expanding hadronic fluid in terms of the Bogoliubov transformation \cite{grib} 
and adiabatic expansion \cite{parker1,parker3}. 
For alternative approaches to cosmological particle creation see, e.g., Refs.\
\cite{hamilton,greenwood}.

The remainder of the paper is organized as follows. 
In Sec.~\ref{chiral} we describe the analog model based on the expanding chiral fluid. 
The cosmological particle creation of pions is studied in Sec.~\ref{creation}  
 in which we derive the spectrum of the created pions and the 
 time dependence of the particle number.
 We estimate the temperature by fitting our spectrum to the 
 Planck black body radiation spectrum.
 In the concluding section,
Sec.~\ref{conclusion}, we summarize our results and
discuss  physical implications.

\section{Expanding chiral fluid}  
\label{chiral}
In this section we describe the hadron fluid in terms of the linear 
sigma model at finite temperature
undergoing a spherically symmetric expansion.
Our model is based on a scalar field Lagrangian with spontaneously broken chiral symmetry
which we describe
in  Sec. \ref{linear}. Then in Sec. \ref{bjorken} we specify the dynamics
of the fluid based on the Bjorken expansion model.
\subsection{Linear sigma model}
\label{linear}
Consider a linear sigma model at finite temperature
in the Minkowski spacetime background.
The background medium is a hadronic fluid  consisting of predominantly pions.
The dynamics of mesons in  such a medium is described by
an effective Lagrangian with spontaneously broken chiral symmetry \cite{bilic2}
\begin{eqnarray}
 {\cal L}_{\rm eff} =
 &&
 \frac{1}{2}(f g^{\mu\nu}
  +g u^{\mu}u^{\nu})(\partial_{\mu}\sigma
\partial_{\nu}\sigma
+\partial_{\mu}\mbox{\boldmath$\pi$}
\partial_{\nu}\mbox{\boldmath$\pi$})
\nonumber\\
&&
-\frac{m_{\sigma}^{2}}{2} \sigma^{2}
- \frac{m_{\pi}^{2}}{2}
\mbox{\boldmath$\pi$}^{2}
-U(\sigma,\mbox{\boldmath$\pi$}),
\label{eq0000}
\end{eqnarray}
 where $g^{\mu\nu}$ is the inverse of the background metric.
As we are working in units in which $\hbar=c=1$ the velocity of the fluid
 $u_{\mu}$ is normalized as $g^{\mu\nu} u_\mu u_\nu=1$.

 The coefficients
 $f$ and $g$ depend  on the local temperature $T$
 and on the parameters of the sigma model:
 the coupling constant $\lambda$ and the pion decay constant $f_\pi$,
 and may be calculated
in perturbation theory.
The scalar  fields 
$\sigma$
and $\pi_i$, where $i=1,2,3$,  represent
fluctuations around the expectation values
 $\langle \sigma\rangle$ and $\langle\pi_i\rangle$,
respectively.
The expectation values of the pion fields  $\langle\pi_i\rangle$ are chosen to vanish always
whereas the expectation value of the sigma field
$\langle \sigma \rangle$, usually referred to as the chiral condensate,  is  temperature dependent
and vanishes at the critical temperature $T_{\rm c}$.
The scaling and universality analysis \cite{son1}
yields  $\langle \sigma \rangle\sim (T-T_{\rm c})^\beta $ 
in the vicinity of the critical point.
At zero temperature $\langle \sigma\rangle $ is normalized to the pion decay constant, i.e., 
$ \langle \sigma \rangle  =
f_{\pi}$ at $T=0$.
The meson masses depend on temperature and 
below the
chiral transition point  are given by
\begin{equation}
m_{\pi}^2 =  0,  \quad
m_{\sigma}^2 = 2\lambda \langle\sigma\rangle^{2}.  
\label{eq43}
\end{equation}
The potential $U$ is
\begin{equation}
 U(\sigma,\mbox{\boldmath$\pi$})=
 \lambda\langle\sigma\rangle \sigma(\sigma^{2}+ \mbox{\boldmath$\pi$}^{2})
+ \frac{\lambda}{4}
 (\sigma^{2}+ \mbox{\boldmath$\pi$}^{2})^2.
\label{eq5}
\end{equation}
The temperature dependence of $\langle \sigma \rangle$
is obtained by
minimizing the thermodynamic potential
$\Omega=-(T/V) \ln Z$
with respect to
 $\langle \sigma \rangle$
at fixed temperature $T$
\cite{bilic1}.
The scaling and universality analysis  \cite{son1}
yields  $\langle \sigma \rangle\sim (T-T_{\rm c})^\beta $ in the vicinity of the critical point 
with $\beta=0.388$ for the O(4) universality class \cite{hasenbusch,toldin}.
Furthermore, the extremum condition was solved numerically at one-loop order
\cite{tolic2,bilic1} and 
the value of the critical temperature $T_{\rm c}= 183$ MeV was found 
with $f_\pi=92.4$ MeV and $m_\sigma=1$ GeV as a phenomenological input.


The action corresponding to the Lagrangian (\ref{eq0000})  may be expressed as  \cite{tolic2}
\begin{eqnarray}
S_{\rm eff} =
&&
\int d^4x \sqrt{-G}\,
\left[\frac{1}{2}G^{\mu\nu}(\partial_{\mu}\sigma
\partial_{\nu}\sigma
+\partial_{\mu}\mbox{\boldmath$\pi$}
\partial_{\nu}\mbox{\boldmath$\pi$})\right.
\nonumber\\
&&
\left. -\frac{c_\pi}{f^2}\left(\frac{m_{\sigma}^{2}}{2} \sigma^{2}
+ \frac{m_{\pi}^{2}}{2}
\mbox{\boldmath$\pi$}^{2}+U(\sigma,\mbox{\boldmath$\pi$})\right) 
\right],
\label{eq0001}
\end{eqnarray}
where the effective metric tensor,
its inverse, and its determinant are given by
\begin{equation}
G_{\mu\nu} =\frac{f}{c_{\pi}}
[g_{\mu\nu}-(1-c_{\pi}^2)u_{\mu}u_{\nu}] ,
\label{eq022}
\end{equation}
\begin{equation}
G^{\mu\nu} =
\frac{c_{\pi}}{f}
\left[g^{\mu\nu}-(1-\frac{1}{c_{\pi}^2})u^{\mu}u^{\nu}
\right],
\label{eq029}
\end{equation}
\begin{equation}
G \equiv \det G_{\mu\nu} = \frac{f^4}{c_{\pi}^2} \det g_{\mu\nu}.
\label{eq030}
\end{equation}
with the pion velocity $c_{\pi}$ defined  by
\begin{equation}
c_{\pi}^2=\frac{f}{f+g} .
\label{eq015}
\end{equation}

It is worth noting that there exists a straightforward map between the metric (\ref{eq022}) and the relativistic 
acoustic metric \cite{moncrief,bilic3,visser2}
\begin{equation}
{\cal G}_{\mu\nu}=\frac{n}{m^2 w c_{\rm s}}
[g_{\mu\nu}-(1-c_{\rm s}^2) u_\mu u_\nu]
\label{eq2109}
\end{equation}
derived for acoustic perturbations in 
an ideal relativistic fluid
 with the adiabatic speed of sound $c_{\rm s}$ defined as
\begin{equation}
 c_{\rm s}^2=\left.\frac{dp}{d\rho} \right|_{s/n} .
 \label{eq2113}
\end{equation}
The symbols $n$, $p$, $\rho$ and $w=(p+\rho)/n$ denote, respectively
the particle number density, pressure, energy density, and
specific enthalpy. The expression (\ref{eq2109}) compared with the original one \cite{bilic3}
differs by a factor  $1/m^2$ which we have introduced here to make the acoustic metric dimensionless.
The mapping between $G_{\mu\nu}$ and ${\cal G}_{\mu\nu}$ is achieved by
identifying the pion velocity $c_\pi$ with the speed of sound $c_{\rm s}$
and the quantity $f$ with $n/(m^2w)$. The physical meaning of $f$ may be seen in the
nonrelativistic limit in which case  
$n\rightarrow \rho/m$ and $w \rightarrow m$, and hence, $f\rightarrow \rho/m^4$.
Thus, in this limit the quantity $f$ is proportional to the  energy density $\rho$.
At zero temperature the energy density is just the rest mass density, i.e.,
$\rho|_{T=0} = m/V= m^4$ so  $f|_{T=0}=1$.

In the absence of the medium (or equivalently at zero temperature) we have $f=1$ and $g=0$.
At nonzero temperature, $f$ and $g$  are derived from the
finite temperature self-energy  $\Sigma(q,T)$ 
in the limit when the external momentum
$q$ approaches zero and
  can be expressed in terms of second derivatives of
 $\Sigma(q,T)$ with respect to $q_0$ and $q_i$.
The  quantities $f$, $g$, and $c_\pi$ as functions of temperature have been calculated
at one-loop level by Pisarski and Tytgat
in the low temperature approximation
\cite{pisarski} 
\begin{eqnarray}
&&
f \simeq 1- \frac{T^2}{12 f_\pi^2}-\frac{\pi^2}{9}\frac{T^4}{f_\pi^2 m_\sigma^2},
\quad g \simeq 1+\frac{8\pi^2}{45}\frac{T^4}{f_\pi^2 m_\sigma^2}, 
\nonumber\\
&&
 c_\pi^2 \simeq 1- \frac{8\pi^2}{45}\frac{T^4}{f_\pi^2 m_\sigma^2} ,
\label{eq2252}
\end{eqnarray}
and by Son and Stephanov for temperatures
close to the chiral transition point \cite{son1,son2} 
(see also Ref.\ \cite{tolic}).
In $d=3$ dimensions one finds 
\begin{eqnarray}
&&
f \simeq c_1 (1-z)^{\nu -2\beta},
\quad f+g \simeq c_2 (1-z)^{-2\beta}, 
\nonumber\\
&&
c_\pi^2 \simeq c_3 (1-z)^\nu ,
\label{eq252}
\end{eqnarray}
in the limit  $z\equiv T/T_{\rm c}\rightarrow 1$. Here  $c_1$, $c_2$, and $c_3$ are constants
and
 $\nu=0.749$ and $\beta=0.388$ are the critical exponents 
 for the O(4) universality class
  \cite{hasenbusch,toldin}. 
Combining this with (\ref{eq2252}) and 
the numerical results at one-loop order \cite{tolic2},
a good fit 
in the entire range $0\leq T \leq T_{\rm c}$ is 
achieved with
\begin{equation}
f =(1-z^2)^{\nu -2\beta}(1-pz^2)^q,
\quad  c_\pi^2 =(1-z^4)^\nu,
\label{eq051}
\end{equation}
where $p$ and $q$ are positive parameters. The constants in (\ref{eq252}) are then fixed to  $c_1=2^{\nu-2\beta}(1-p)^q$,
 $c_2=2^{-\nu-2\beta}(1-p)^q$ and $c_3=2^{2\nu}$.
With (\ref{eq051}) we capture the main features: with the values $p=0.1$ and $q=3.58$ we  match 
the zero temperature limit \cite{pisarski}
 and we recover the correct critical behavior (\ref{eq252}).

The variation of the action (\ref{eq0001}) yields  
the Klein--Gordon wave equation in curved space
\begin{equation}
\frac{1}{\sqrt{-G}}\,
\partial_{\mu}
(\sqrt{-G}\,
G^{\mu\nu})
\partial_{\nu}\varphi
+\frac{c_{\pi}}{f^2}\left[m_\varphi^2 +V_\varphi(\sigma,
\mbox{\boldmath{$\pi$}})\right]\varphi
=0 ,
\label{eq028}
\end{equation}
where $\varphi$ stands for $\pi_i$ or $\sigma$ and 
\begin{equation}
V_\varphi(\sigma,
\mbox{\boldmath{$\pi$}})=2\frac{\partial U}{\partial \varphi^2}
\label{eq214}
\end{equation}
are the corresponding interaction potentials.

\subsection{Spherical Bjorken expansion}
\label{bjorken}
In this section we will completely specify the analog metric $G_{\mu\nu}$
by fixing  
the background metric $g_{\mu\nu}$ and the velocity field $u_\mu$ and
by deriving the spacetime dependence of the quantities $f$ and $c_\pi$.
So far, $f$ and $c_\pi$ are expressed  as functions of
temperature via (\ref{eq051}) and we will see 
that once we specify the dynamics of the fluid, the spacetime dependence 
of $f$ and $c_\pi$ will
follow from
their temperature dependence. 
 
To this end we consider a boost invariant 
Bjorken-type spherical expansion \cite{bjorken}.
This type of expansion has been recently applied to mimic
an open FRW metric 
 in a relativistic BE system \cite{fagnocchi,tolic3}.
In this model the radial three-velocity in
 radial coordinates
$x^\mu=(t,r,\theta,\phi)$ is  a simple function 
$v=r/t$. Then the four-velocity  
 is  given by
\begin{equation}
u^\mu= (t/\tau, r/\tau,0,0) ,
\label{eq144}
\end{equation}
where $\tau=\sqrt{t^2-r^2}$ is the 
{\em proper time} of observers comoving along the fluid worldlines.
With the substitution
\begin{eqnarray}
& &t=\tau \cosh y ,
\nonumber \\
& & r=\tau \sinh y ,
\label{eq147}
\end{eqnarray}
the four-velocity  velocity is expressed as
\begin{equation}
u^\mu=(\cosh y,\sinh y,0, 0).
\label{eq146}
\end{equation}
The substitution (\ref{eq147}) may be regarded as a coordinate transformation
from ordinary radial coordinates 
to  new coordinates
$(\tau,y,\theta,\phi)$
in which
the flat background metric
takes the form
\begin{equation}
g_{\mu\nu}= {\rm diag} \left(1,  -\tau^2, -  
\tau^2\sinh^2\! y, - \tau^2\sinh^2\! y \sin^2\!\theta \right).
\label{eq218}
\end{equation}
Thus, the transformation (\ref{eq147}) maps the spatially flat
Minkowski spacetime into an expanding FRW spacetime with cosmological scale $a=\tau$ and negative spatial 
curvature. The resulting flat spacetime with metric (\ref{eq218}) is  known in cosmology 
as the {\it Milne universe} \cite{milne}.

The velocity components in this coordinate frame are  $u^\mu=(1,0,0,0)$,
and hence, the new coordinate frame is comoving.
Using this and  (\ref{eq218}) 
from (\ref{eq022}) we obtain 
the analog metric in a diagonal form
\begin{equation}
G_{\mu\nu}
 =
\left(\begin{array}{cccc}
  b^2  &          &  &   \\
           & -a^2  &     &     \\
           &          & -a^2\sinh^2\! y &   \\
           &          &                  & -a^2\sinh^2\! y \sin^2 \theta 
\end{array} \right) ,
\label{eq243}
\end{equation}
where 
\begin{equation}
b=\sqrt{fc_\pi},  \quad a=\tau\sqrt{\frac{f}{c_\pi}} .
\label{eq108}
\end{equation}
This metric is of the form of a FRW spacetimes with negative
spatial curvature provided the quantities $f$ and ${c_\pi}$ are
functions of time only.

We could, in principle, convert the metric (\ref{eq243}) back to the original coordinates
$(t,r,\theta,\phi)$ using the inverse of the transformation (\ref{eq147})
and proceed with calculations in the laboratory frame.
However, for our purpose it is advantageous to do the calculations
in the comoving reference frame mainly for the following reasons. 
The original coordinate frame (or laboratory frame) 
is not suitable for thermodynamic considerations since the thermodynamic variables, such as temperature,
are always defined in the fluid rest frame or the comoving frame. 
Besides, as we shall shortly demonstrate,  the comoving reference frame yields 
an analog FRW cosmology.
Hence, from now on we work in the comoving reference frame referring to  
the proper time $\tau$ simply as the {\em time}.

To find the time dependence of $f$ and ${c_\pi}$
we shall use
(\ref{eq051}) and the fact that 
the temperature of the expanding chiral fluid is, to a good approximation,  proportional to $\tau^{-1}$.
This follows from the fact that the expanding hadronic matter is dominated by massless pions,
and hence, the density and pressure of the fluid 
may be approximated by $\rho=(\pi^2/10) T^4$ and $p=\rho/3$ for an ideal massless pion gas 
\cite{landau}.
Using this and the continuity equation 
 %
$u^\mu \rho_{;\mu}+(p+\rho){u^\mu}_{;\mu}=0$,
where the subscript $;\mu$ denotes the covariant derivative  associated with
the background metric (\ref{eq218}),
  one finds
 
\begin{equation}
\frac{T}{T_{\rm c}} = \frac{\tau_{\rm c}}{\tau} .
\label{eq007}
\end{equation}
Here $T_{\rm c}$ is the critical temperature of the chiral transition
and $\tau_{\rm c}$ may be fixed from 
the phenomenology of high energy collisions.
For example, if we take $T_{\rm c}=0.183$ GeV,
then a typical value of
$\rho=1$ GeV/fm$^3$ at  $\tau\approx 5$ fm 
\cite{kolb-russkikh}
is obtained with 
$\tau_{\rm c}\approx 8.2 \; {\rm fm} =41.6\; {\rm GeV}^{-1}$. 
 The physical range of $\tau$ is fixed by Eq.~(\ref{eq007}) since the available temperature ranges
between $T=0$ and $T=T_{\rm c}$. Hence, the time range is
$\tau_{\rm c}\leq \tau < \infty$. 
In the following we 
keep $\tau_{\rm c}$ unspecified so that physical quantities of the dimension of time or length are 
expressed in units of $\tau_{\rm c}$.

Thus, the quantities $f$ and $c_\pi$ being temperature dependent
are implicit functions of $\tau$ 
through the time dependent $T=T(\tau)$. 
In this way, the metric (\ref{eq243}) falls into the class of FRW spacetimes with negative
spatial curvature.
More explicitly, using (\ref{eq051}) and (\ref{eq007}) we have
\begin{equation}
 a=\tau \left(1-z^2\right)^{\nu/4-\beta}(1+z^2)^{-\nu/4}(1-pz^2)^{q/2},
  \label{eq2108}
 \end{equation}
\begin{equation}
 b=\left(1-z^2\right)^{3\nu/4-\beta}(1+z^2)^{\nu/4}(1-pz^2)^{q/2},
  \label{eq2208}
 \end{equation} 
   where $z=\tau_{\rm c}/\tau=T/T_{\rm c}$, $\nu=0.749$, $\beta=0.388$,
   $p=0.1$, and $q=3.58$.

In the next section we will have to address 
the criteria for distinguishing an adiabatic from a sudden  regime.
The relevant scale  for these criteria 
is set by the Hubble parameter which for this cosmology is defined as
\begin{equation}
 H =\frac13 \nabla_\mu u^\mu=\frac{1}{ab}\frac{\partial a}{\partial \tau},
 \end{equation}
 where $\nabla_\mu$ denotes the covariant derivative associated with the metric
 (\ref{eq243}).
For large $\tau$ the quantity $H$ goes to zero  
as $H\sim 1/\tau$  and near the critical point
diverges as
\begin{equation}
H\simeq - 0.2 \, \tau^{-2} (1-1/\tau)^{\beta -3\nu/4-1},
  \label{eq2308}
 \end{equation}
 where $\tau$ is measured in units of $\tau_{\rm c}$ and $H$ in units of 
$\tau_{\rm c}^{-1}$.

\section{Creation of pions in analog cosmology}
\label{creation}
The particles associated with quantum fluctuations of the chiral field
are pions and sigma mesons. Since the chiral fluid is expanding and
the particles experience an effective time-dependent metric, 
the pions and sigma mesons will be created during the expansion
in complete analogy with the standard cosmological
particle creation.
As a consequence, the  pion and sigma-mesons numbers will not be
conserved and a vacuum state generally evolves into a multiparticle
state. 
In Sec.\ \ref{canonical} we review the standard procedure of canonical quantization of
scalar fields in a FRW geometry and the derivation of Bogoliubov coefficients.
In Sec.\ \ref{particle} we deal with the particle interpretation  ambiguity 
in a time-dependent geometry. Next, we solve the Klein--Gordon equation 
with the help of the WKB ansatz and in Sec.\ \ref{results} we present the numerical results.

\subsection{Canonical quantization} 
\label{canonical}
The effective action (\ref{eq0001}) with (\ref{eq5}) is of the $\varphi^4$ type.
However, as we are primarily interested in the effects of cosmological particle creation
we shall in the following disregard the self-interaction terms in the potential.
In other words,  we do not consider particle production caused by  the self-interaction
although this effect may be significant \cite{birrell}.
Hence,  we can split the action (\ref{eq0001}) into a  sum 
of the actions for each field: 
\begin{equation}
S =
\frac{1}{2}\int d^4x \sqrt{-G}\,
\left[G^{\mu\nu}\partial_{\mu}\varphi
\partial_{\nu}\varphi
-m_{\rm eff}^2 \varphi^2 
-\xi\varphi^2  R\right],
\label{eq2001}
\end{equation}
where $\varphi$ stands for $\pi_i$ or $\sigma$.
Here
we have introduced a time-dependent effective mass defined by
\begin{equation}
m_{\rm eff}^2=\frac{c_{\pi}}{f^2}m^2=\frac{\tau^3}{a^3b}m^2 ,
\label{eq2103}
\end{equation}
where
$m$ stands for $m_\pi$ or $m_\sigma$ as given by 
(\ref{eq43}).
For completeness, we have included 
the nonminimal coupling term  of the scalar fields to the  
effective scalar curvature $R$. This term is required by renormalization in curved spacetime because, 
even if the renormalized  $\xi=0$,
 loop corrections would induce a nonminimal coupling term of this type
\cite{birrell,parker2}.
As we shall see, this term introduces another type of criticality
in addition to the critical behavior owing to the above-mentioned 
chiral symmetry breaking and restoration at finite temperature.
The  field equation derived from (\ref{eq2001}) is the free Klein--Gordon equation 
\begin{equation}
\frac{1}{\sqrt{-G}}\,
\partial_{\mu}
(\sqrt{-G}\,
G^{\mu\nu})
\partial_{\nu}
\varphi
+m_{\rm eff}^2\varphi
+\xi R\varphi
=0 
\label{eq2002}
\end{equation}
in curved spacetime with metric (\ref{eq243}).
\subsubsection{Schr\"{o}dinger representation}
In the canonical quantization formalism we expand the field $\varphi$ 
\begin{equation}
\varphi(x) =
\sum_J \left[
a_J \varphi_J(x)+a_J^\dag  \varphi_J(x)^* 
\right] ,
\label{eq2003}
\end{equation}
where the time independent particle creation and annihilation operators 
(the ``Schr\"{o}dinger picture") satisfy
the commutation relations
\begin{equation}
[a_J,a_K^\dag]=\delta_{JK}
 \label{2004}
\end{equation}
and $\varphi_J(x)$ are solutions to  (\ref{eq2002})
labeled by a collective index $J$.
In spherical coordinates 
$J\equiv\{k,l,m\}$, where $l$ is the angular momentum, $m$ is its projection, 
and $k$ is the magnitude of the comoving momentum related to the physical momentum as $p=k/a$.
Note that in the coordinate frame $(\tau,y,\theta,\phi)$ with metric (\ref{eq243})
$k$ is dimensionless.
Thus, the physical energy of a particle is 
\begin{equation}
 E=\sqrt{k^2/a^2+m_{\rm eff}^2}, 
 \label{eq2104}
\end{equation}
where $m_{\rm eff}$ is defined by (\ref{eq2103}).

In a  hyperbolic space,  the large volume limit $V\rightarrow \infty$
can be applied, and
the sum over discrete momentum is replaced by an integral over continuous $k$.
Then the sum over $J$ in (\ref{eq2003}) 
can be written as \cite{grib}
\begin{equation}
\sum_J = \int_0^\infty dk\sum_{l=0}^\infty \sum_{m=-l}^l.
\label{eq2005}
\end{equation}
As usual, we can separate the time and space dependence using
\begin{equation}
\varphi_J(x)=\left(\frac{b}{2a^3}\right)^{1/2}\chi_k(\tau)\Phi_J(\mbox{\boldmath{$x$}}).
\label{eq2006}
\end{equation}
Then, the functions $\chi_k$ and $\Phi_J$  satisfy
\begin{equation}
\chi_k''+\Omega^2(\tau)\chi_k=0,
\label{eq2007}
\end{equation}
\begin{equation}
\Delta \Phi_J+k^2\Phi_J=0,
\label{eq2008}
\end{equation}
respectively.
Here and from here on, the prime $'$ denotes a partial derivative with respect to $\tau$.

The differential operator $\Delta$ is the Laplace--Beltrami operator
on the three-dimensional space with line element 
\begin{equation}
dl^2=\gamma_{ij} dx^i dx^j=dy^2+\sinh^2 y (d\theta^2+\sin^2\theta d\phi^2).
 \label{eq2009}
\end{equation}
The metric-dependent factor on the right-hand side of (\ref{eq2006})
is introduced in order to get rid of the first-order derivative in the equation for $\chi$. 
The time-dependent function $\Omega$ is given by
\begin{equation}
\Omega^2=\omega^2+\left(\xi - \frac16 \right)b^2R +\Sigma ,
\label{eq2010}
\end{equation}
where
\begin{equation}
\Sigma=
\frac14\left(\frac{b}{a}\right)^2
\left(\frac{a}{b}\right)'^{\, 2}-\frac12\left(\frac{b}{a}\right)
\left(\frac{a}{b}\right)'' ,
\label{eq2120}
\end{equation}
$R$ is the Ricci scalar,
   \begin{equation}
 R=\frac{6}{a^2} \left[\frac{a a''}{b^2}
 +\frac{a'}{b}\left(\frac{a}{b}\right)'-1\right],
 \label{eq2110}
 \end{equation} 
and
\begin{equation}
\omega^2=b^2 E^2=\frac{b^2}{a^2}k^2+ \frac{\tau^3 b}{a^3}m^2,
\label{eq2011}
\end{equation}
with mass $m=m(\tau)$ generally depending on $\tau$ by way of its temperature dependence. 
The spacetime defined by the metric 
(\ref{eq243}) with (\ref{eq2108}) and (\ref{eq2208}) 
has a curvature singularity since the Ricci scalar
diverges at $\tau_{\rm c}$
as 
\begin{equation}
R\sim (\tau-\tau_{\rm c})^{-2}.
 \label{eq2111}
\end{equation}


The solutions to (\ref{eq2008}) are known and
the explicit form of $\Phi_J(\mbox{\boldmath{$x$}})$ may be found in 
Ref.\ \cite{grib}. Here we  only use the following properties
\begin{equation}
\int d^3x \sqrt{\det \gamma}\, \Phi_J^*(\mbox{\boldmath{$x$}})\Phi_K(\mbox{\boldmath{$x$}})=\delta_{JK},
 \label{2101}
\end{equation}
\begin{equation}
\sum_{lm} |\Phi_J(\mbox{\boldmath{$x$}})|^2 
=\frac{k^2}{2\pi^2}.
 \label{2103}
\end{equation}
Using these properties, the sum over $l$ and $m$ in (\ref{eq2005}) may be carried out in the case of spherical symmetry.
First, using (\ref{2101}) with $J=K$  we rewrite (\ref{eq2005}) as
\begin{equation}
\sum_J=\int_0^\infty dk \sum_{lm} \int d^3x \sqrt{\det \gamma}\, \Phi_J^*(\mbox{\boldmath{$x$}})\Phi_J(\mbox{\boldmath{$x$}}).
 \label{2104}
\end{equation}
Then, by (\ref{2103}) we obtain
\begin{equation}
\sum_J=\frac{V}{2\pi^2}\int_0^\infty dk k^2 ,
 \label{2105}
\end{equation}
where $V$ denotes the comoving proper volume
\begin{equation}
V=  \int d^3x \sqrt{\det\gamma}\,  .
 \label{2106}
\end{equation}

To solve (\ref{eq2007}) we first impose the condition
\begin{equation}
\chi_k\chi_k^{*\,\prime} -\chi_k^* \chi_k'=2i , 
\label{eq2012}
\end{equation}
which we can do because the left-hand side, the Wronskian, is a constant of motion of
(\ref{eq2007}).
Next we make use of the WKB ansatz which automatically meets the condition (\ref{eq2012}),
\begin{equation}
\chi_k(\tau) = W(\tau)^{-1/2}e^{-i\int d\tau' W(\tau')},
\label{eq2013}
\end{equation}
  where
the positive function $W(\tau)$ satisfies 
\begin{equation}
W^2=\Omega^2+ W^{1/2}(W^{-1/2})''.
\label{eq2014}
\end{equation}
A solution to (\ref{eq2014}) may be expressed as 
\begin{equation}
W(\tau)=\lim_{n\rightarrow\infty} W^{(n)}(\tau),
\label{eq2015}
\end{equation}
where the series $\{W^{(n)}\}$ is obtained from the adiabatic expansion
\cite{parker1}
\begin{equation}
W^{(n)}(\tau)=\sum_{i=0}^n \omega^{(i)}(\tau).
\label{eq2016}
\end{equation}
The contribution $\omega^{(i)}(\tau)$
containing derivatives of $a$, $b$, and $\omega$ with respect to $\tau$ of order $i$ 
may be found iteratively at each adiabatic order $i$ starting with $\omega^{(0)}(\tau)=\omega(\tau)$.
The next nonvanishing term in the series in (\ref{eq2016}) is of the order $i=2$ and is given by
\begin{equation}
\omega^{(2)}=\left(\xi - \frac16 \right)\frac{b^2R}{2\omega} +\frac{\Sigma}{2\omega}
+\frac{(\omega^{-1/2})''}{\omega^{1/2}} .
\label{eq2031}
\end{equation}

Using (\ref{eq2016}) we define the adiabatic modes of order $n$ as
\begin{equation}
\varphi^{(n)}_J(\tau,\mbox{\boldmath{$x$}})=\left(\frac{b}{2a^3}\right)^{1/2}\chi_k^{(n)}(\tau)\Phi_J(\mbox{\boldmath{$x$}}),
\label{eq2019}
\end{equation}
where
\begin{equation}
\chi_k^{(n)}(\tau) = \left(W^{(n)}(\tau)\right)^{-1/2}e^{-i\int d\tau' W^{(n)}(\tau')} .
\label{eq2020}
\end{equation}
\subsubsection{Heisenberg representation}
In the following we assume the adiabatic invariance of the particle number in each mode of the field
$\varphi$, i.e., we require that the particle number in each mode should
be constant \cite{parker4} in the limit of an infinitely slow expansion. 
To meet this requirement we expand the field operator $\varphi(x)$ and its derivative $\varphi'(x)$ in terms of 
the time-dependent operators (the ``Heisenberg picture'') $a_J (\tau)$ 
\begin{equation}
\varphi(x) =
\sum_J \left[
a_J(\tau) \tilde{\varphi}_J(\tau,\mbox{\boldmath{$x$}})+a_J^\dag(\tau)  \tilde{\varphi}_J(\tau,\mbox{\boldmath{$x$}})^*
\right],
\label{eq2017}
\end{equation}
\begin{equation}
\varphi'(x) =
 \sum_J\left[
a_J(\tau) \tilde{\varphi}_J^{\prime}(\tau,\mbox{\boldmath{$x$}})+a_J^\dag(\tau)  \tilde{\varphi}_J^{\prime}(\tau,\mbox{\boldmath{$x$}})^*
\right],
\label{eq2018}
\end{equation}
where 
\begin{equation}
\tilde{\varphi}_J(\tau,\mbox{\boldmath{$x$}})=\left(\frac{b}{2a^3}\right)^{1/2}\tilde{\chi}_k(\tau)\Phi_J(\mbox{\boldmath{$x$}}),
\label{eq2106}
\end{equation}
\begin{equation}
\tilde{\chi}_k(\tau) = \left(\tilde{W}(\tau)\right)^{-1/2}e^{-i\int d\tau' \tilde{W}(\tau')} ,
\label{eq2220}
\end{equation}
and $\tilde{W}(\tau)$ are  conveniently chosen smooth functions of $\omega$, $a$, and $b$, 
and their derivatives, such that
\begin{equation}
\tilde{W}(\tau_0) =W(\tau_0), \quad \tilde{W}'(\tau_0) =W'(\tau_0)
\end{equation}
at some initial $\tau=\tau_0$. In addition, for a given $k$ the function
$\tilde{W}(\tau)$ should coincide with $W^{(0)}(\tau)$  at lowest adiabatic order, i.e.,
we require $\tilde{W}(\tau)=\omega(\tau)$
at lowest adiabatic order.
The function $\tilde{W}(\tau)$ is, obviously, not unique but the adiabatic expansion 
(\ref{eq2016}) offers 
a natural choice--each adiabatic mode  $W^{(n)}(\tau)$ 
of order $n$ satisfies the above criteria.
However,  the lowest-order mode $W^{(0)}$ is unacceptable as it leads to UV 
divergence in particle production rate \cite{fulling,landete}. 
We shall shortly discuss this point in more detail.

 The simplest, and perhaps the most natural choice
is $\tilde{W}(\tau)\equiv W^{(1)}(\tau)=\omega(\tau)$.  
Another choice, $\tilde{W}(\tau)\equiv\Omega(\tau)$, which meets the above criteria 
also seems natural since $\Omega$ appears in the harmonic oscillator equation
 (\ref{eq2007}) as the (time-dependent)
frequency. However,  in contrast to $W^{(1)}(\tau)$, the function
$\Omega(\tau)$ contains  derivative terms of second order but not all such terms that appear in
the next order adiabatic mode $W^{(2)} (\tau)$. Hence, the choice $\tilde{W}(\tau)\equiv\Omega(\tau)$
is incomplete and inconsistent from the adiabatic expansion point of view.
Another problem with $\tilde{W}(\tau)\equiv\Omega(\tau)$, as we shall shortly demonstrate,  is that
it yields a nonvanishing particle creation rate in the conformal case.

\subsubsection{Bogoliubov transformation}

 The time-dependent operators $a_J(\tau)$ and $a_J^\dag(\tau)$ are related to $a_J(\tau_0)\equiv a_J$ and 
 $a_J^\dag(\tau_0)\equiv a_J^\dag$ 
 via the Bogoliubov transformation
\cite{grib,pavlov}
\begin{equation}
 a_J(\tau)=\alpha_k(\tau)a_J +\beta_k^{\, *}(\tau) a_{\bar{J}}^\dag \vartheta_J ,
 \label{eq2022}
\end{equation}
\begin{equation}
 a_J^\dag(\tau)=\alpha_k^{\, *}(\tau)a_J^\dag +\beta_k(\tau) a_{\bar{J}} \vartheta_J^* ,
 \label{eq2023}
\end{equation}
where  
the coefficients satisfy
\begin{equation}
 \alpha_k(\tau_0)=1, \quad \beta_k(\tau_0)=0, \quad
 |\alpha_k|^2-|\beta_k|^2=1.
 \label{eq2024}
\end{equation}
The conjugate label $\bar{J}$ is defined so that
\begin{equation}
 \Phi_J^*(\mbox{\boldmath{$x$}})=\vartheta_J \Phi_{\bar{J}} (\mbox{\boldmath{$x$}})
 \label{eq2025}
\end{equation}
and $\vartheta_J$ is a phase with property
\begin{equation}
 \vartheta_{\bar{J}}=\vartheta_J^* .
 \label{eq2026}
\end{equation}
For example, if $J\equiv \{k,l,m\}$, we have $\bar{J}\equiv \{k,l,-m\}$ and 
$\vartheta_J=(-1)^m$.

Consistency of the expansions  (\ref{eq2017}) and (\ref{eq2018}) with (\ref{eq2003}) implies
a relationship between
exact solutions $\chi_k(\tau)$ and the known functions $\tilde{\chi}_k(\tau)$.
Plugging (\ref{eq2022}) and (\ref{eq2023}) into (\ref{eq2017}) and (\ref{eq2018}), 
and comparing with (\ref{eq2003}) one finds 
\begin{equation}
 \chi_k=\alpha_k\tilde{\chi}_k +\beta_k\tilde{\chi}_k^{\, *} ,
 \label{eq2027}
\end{equation}
\begin{equation}
 \chi_k'=\alpha_k\tilde{\chi}_k^{\, \prime} +\beta_k\tilde{\chi}_k^{\, *\, \prime} .
 \label{eq2028}
\end{equation}
 In obtaining these equations we have used
 (\ref{eq2025}) and (\ref{eq2026}). 
  Clearly, by virtue of (\ref{eq2024})  we have 
\begin{equation}
\chi_k(\tau_0)=\tilde{\chi}_k(\tau_0), \quad \chi_k'(\tau_0)=\tilde{\chi}_k^{\,\prime}(\tau_0),
\label{eq2021}
\end{equation}
which serve as initial conditions when solving equation (\ref{eq2007}).
From (\ref{eq2027}) and (\ref{eq2028}) we find the explicit expressions for the Bogoliubov coefficients:
\begin{equation}
 \alpha_k= \frac{1}{2i}\left(\chi_k\tilde{\chi}_k^{\, *\, \prime}  -\tilde{\chi}_k^*\chi_k'\right),
 \label{eq2128}
\end{equation}
 \begin{equation}
 \beta_k= \frac{1}{2i}\left(\tilde{\chi}_k\chi_k'  -\chi_k\tilde{\chi}_k^{\,\prime}\right).
 \label{eq2129}
\end{equation}

\subsection{Particle interpretation}
\label{particle}
As is well known, there exists an intrinsic ambiguity of the particle interpretation
in spacetimes with a time-dependent metric, in particular in a FRW spacetime \cite{parker3}.
In this section, we present a simple demonstration of this ambiguity and the prescription how to remove it.

At $\tau_0$, 
the vacuum state vector $|\;\rangle$ is defined as the state which is annihilated by 
 the operator $a_J$, i.e., $a_J|\;\rangle=0$. A one-particle state with quantum numbers $J$
 is defined using the creation operator 
 $a_J^\dag$ acting on the vacuum, i.e.,  $a_J^\dag|\;\rangle=|J\rangle$, so that
 in the coordinate representation we define
 \begin{equation}
 \langle \mbox{\boldmath{$x$}}|J\rangle= \varphi_J(\tau_0,\mbox{\boldmath{$x$}}).
 \label{eq3001}
 \end{equation}
In the Heisenberg picture the state $|\;\rangle$ is time independent whereas $a_J(\tau)$ and 
$a_J^\dag(\tau)$
evolve with $\tau$ according to (\ref{eq2022}) and (\ref{eq2023}) so
\begin{equation}
 a_J(\tau)|\;\rangle\neq 0
\end{equation}
for $\tau\neq\tau_0$. 

In the Schr\"odinger picture, the vacuum state $|\;\rangle$ evolves  into
a new state vector $|\;\rangle_\tau$ which represents the vacuum with respect to $a_J(\tau)$ such that
\begin{equation}
 a_J(\tau)|\;\rangle_\tau = 0,
\quad
 a_J^\dag(\tau)|\;\rangle_\tau=|J\rangle_\tau ,
 \label{eq2229}
\end{equation}
where the one-particle state in the coordinate representation is defined as
\begin{equation}
 \langle \mbox{\boldmath{$x$}}|J\rangle_\tau= \tilde{\varphi_J}(\tau,\mbox{\boldmath{$x$}}).
 \end{equation} 
  From (\ref{eq2022}) and (\ref{eq2229}), it follows that
 \begin{equation}
 a_J|\;\rangle_\tau\neq 0 ,
\end{equation}
so the state vector  $|\;\rangle_\tau$ at late times is different from
the state vector $|\;\rangle$ containing no particles at an early time $\tau_0$, 
and hence, there is no unambiguous unique Heisenberg state which can be identified as
the vacuum state.

The total  number of particles with quantum number $J$ created at time $\tau$ is
\begin{equation}
 N_J(\tau) \equiv\, _\tau\langle\;|a_J^\dag a_J|\;\rangle_\tau=\langle\;|a_J^\dag(\tau) a_J(\tau)|\;\rangle
 =|\beta_k(\tau)|^2 ,
 \label{eq2029}
\end{equation}
where the last two equations follow from (\ref{eq2022}) and (\ref{eq2023}).
Then, the particle number density is the total number divided by the physical volume $a^3V$, i.e.,
\begin{equation}
 n(\tau) =\frac{1}{a^3V}\sum_J N_J(\tau) 
 =\frac{1}{2\pi^2a^3}\int_0^\infty dk k^2|\beta_k(\tau)|^2,
 \label{eq2030}
\end{equation}
where  we have exploited spherical symmetry and used (\ref{2106}) to replace the sum by an integral.
Thus, the occupation number of created particles is equal to the square of the magnitude of the 
Bogoliubov coefficient $\beta_k$.

The square of the Bogoliubov coefficient may be  obtained directly from (\ref{eq2129}), and by way of 
(\ref{eq2013}) and (\ref{eq2220})
 may be conveniently expressed in terms of $W$ and $\tilde{W}$:
\begin{equation}
 |\beta_k|^2= \frac14 \left[\frac{\tilde{W}}{W}+\frac{W}{\tilde{W}}
 +\frac{1}{4\tilde{W}W}\left( \frac{W'}{W}-\frac{\tilde{W}^{\, \prime}}{\tilde{W}}\right)^2 -2\right].
 \label{eq2032}
\end{equation}
The above-mentioned ambiguity is in the choice of $\tilde{\chi}(\tau)$ or $\tilde{W}(\tau)$
which one has to fix in order to evaluate the right-hand side of (\ref{eq2032}).
 As we discussed previously, a natural choice would be an adiabatic mode $\chi^{(n)}$.
 In this case, to maintain consistency with the adiabatic expansion,
 we  must keep
only the terms up to the adiabatic order $n$ \cite{landete} in the derivatives  
$\tilde{W}'=W^{(n)\, \prime}$ in (\ref{eq2032}). 
For example, consider $n=0$ and $n=1$. 
It turns out that $\omega^{(1)}=0$ in the adiabatic expansion (\ref{eq2016}) so
$W^{(1)}=W^{(0)}=\omega$ and $W^{(1)\, \prime}=\omega^\prime$, but $W^{(0)\, \prime}$ 
must be set to zero.

Furthermore, according to the adiabatic regularization prescription, the choice of $n$ is dictated by
the asymptotic UV behavior of the integrand  in (\ref{eq2030}): 
one must use the minimal $n$ such that
$|\beta_k|\rightarrow 0$ faster than $k^{-3/2}$ as $k\rightarrow \infty$ \cite{landete}.
For a FRW  metric of the general form (\ref{eq243}) the integral (\ref{eq2030}) is UV divergent for $n=0$
apart from some special cases discussed in detail by Fulling \cite{fulling}.
For example, it is easy to verify that such a special case is realized if $b=a$ in (\ref{eq243}),
which corresponds to the conformal form of a spatially hyperbolic metric. 
In a more general case, i.e., if $b \neq a$
the integral (\ref{eq2030}) is UV divergent for $n=0$
and converges for $n\geq 1$.
We shall therefore work with $n=1$ and check the convergence explicitly.
Hence, the final expression
which will be used in our numerical calculations is
\begin{equation}
 |\beta_k|^2= \frac14 \left[\frac{\omega}{W}+\frac{W}{\omega}
 +\frac{1}{4\omega W}\left( \frac{W'}{W}-\frac{\omega'}{\omega}\right)^2 -2\right] ,
 \label{eq2033}
\end{equation}
where $\omega(\tau)$ is given by (\ref{eq2011}) and $W(\tau)$ is a solution to (\ref{eq2014}) that
satisfies the initial conditions
\begin{equation}
 W(\tau_0)=\omega(\tau_0), \quad W'(\tau_0)=\omega'(\tau_0),
 \label{eq2034}
\end{equation}
at conveniently chosen $\tau=\tau_0$.

To check the UV limit
we will make  use of the asymptotic expressions for $W$.
From (\ref{eq2016}) and (\ref{eq2031}) 
in the limit
 $k\rightarrow \infty$
 we find
\begin{equation}
W=\omega+\omega^{(2)}+{\cal O}(\omega^{-3}).
\label{eq2039}
\end{equation}
\begin{equation}
\omega^{(2)}=\left(\xi - \frac16 \right)\frac{b^2R}{2\omega}+{\cal O}(\omega^{-3}),
\label{eq2038}
\end{equation}
Then, applying (\ref{eq2033}) we obtain
\begin{equation}
|\beta_k|^2=\frac{1}{32} \left(\xi - \frac16 \right)^2\frac{b^4R^2}{\omega^4}+{\cal O}(\omega^{-6}).
\label{eq2040}
\end{equation}
Hence, the integrand in (\ref{eq2030})
converges as $1/k^2$ and the particle number has a regular UV limit
as expected. The result (\ref{eq2040}) is generally valid for particle creation in 
any FRW universe with a  metric of the form (\ref{eq243}).

\begin{figure*}[t]
\begin{center}
\includegraphics[width=0.45\textwidth ]{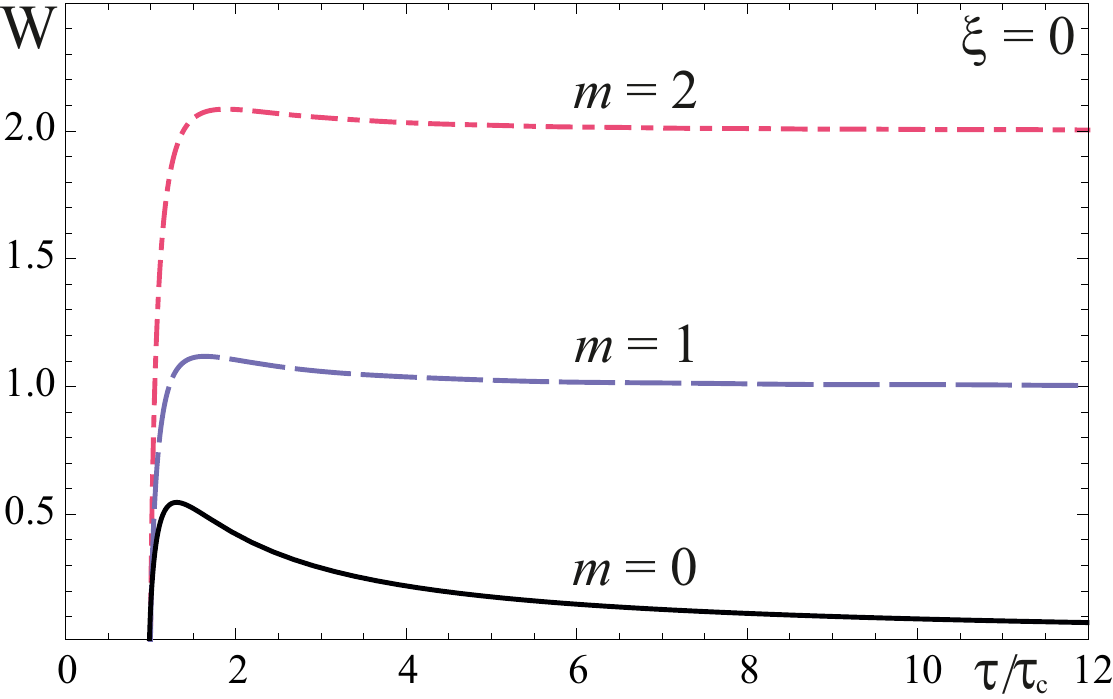}
\hspace{0.02\textwidth}
\includegraphics[width=0.45\textwidth]{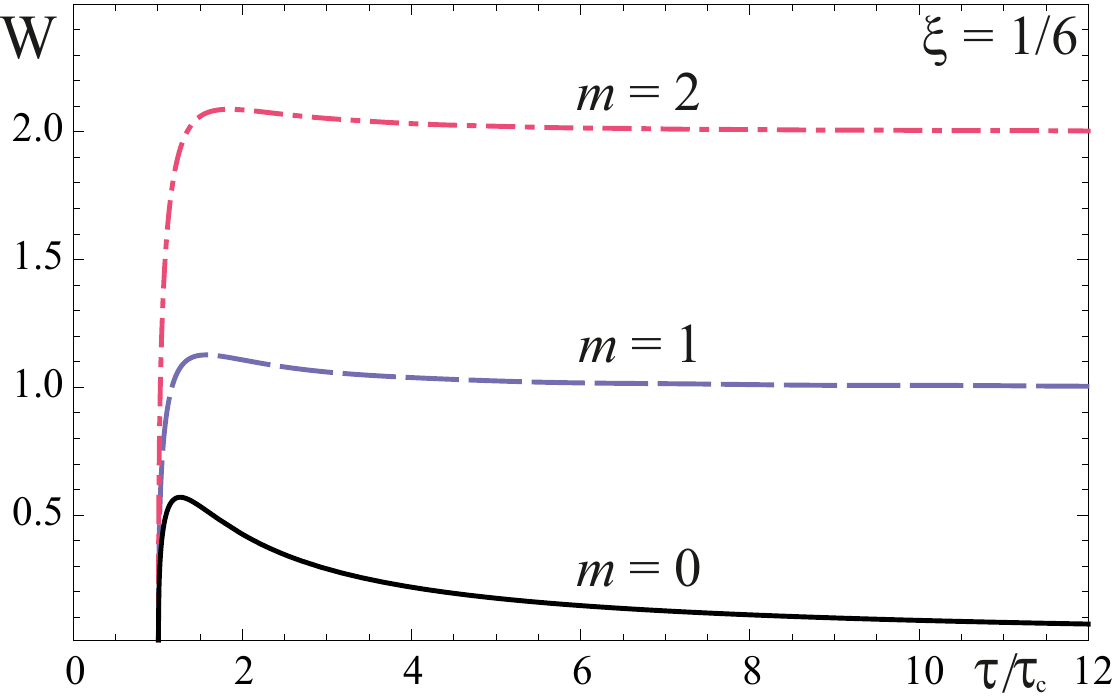}
\\
\vspace{0.02\textwidth}
\includegraphics[width=0.45\textwidth]{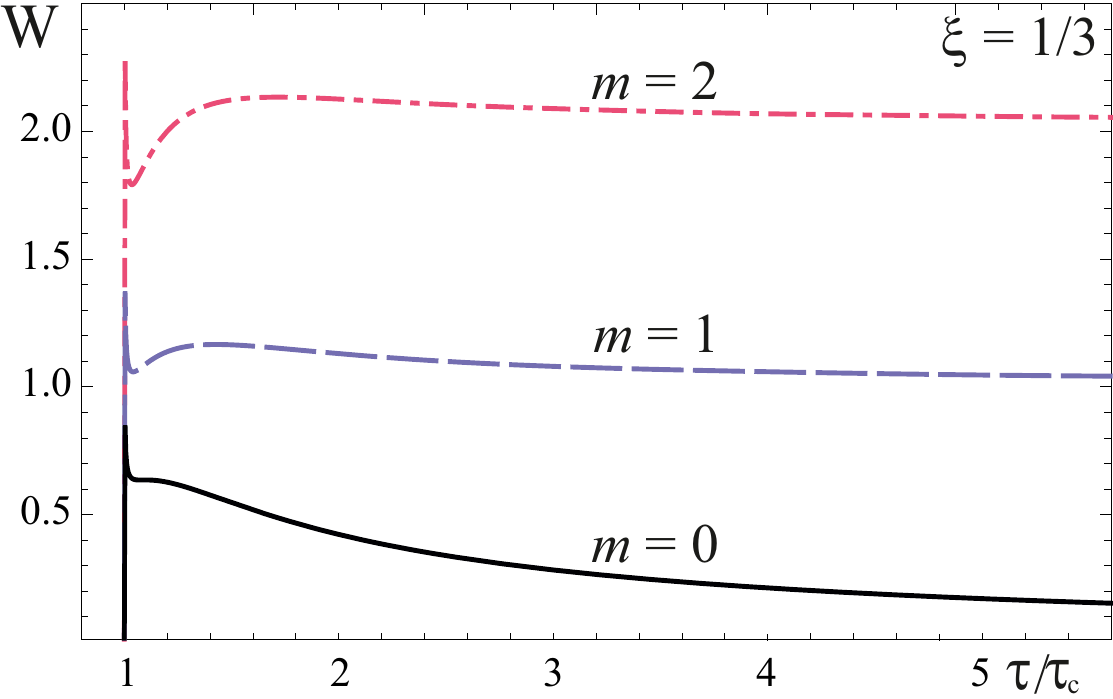}
\hspace{0.02\textwidth}
\includegraphics[width=0.45\textwidth]{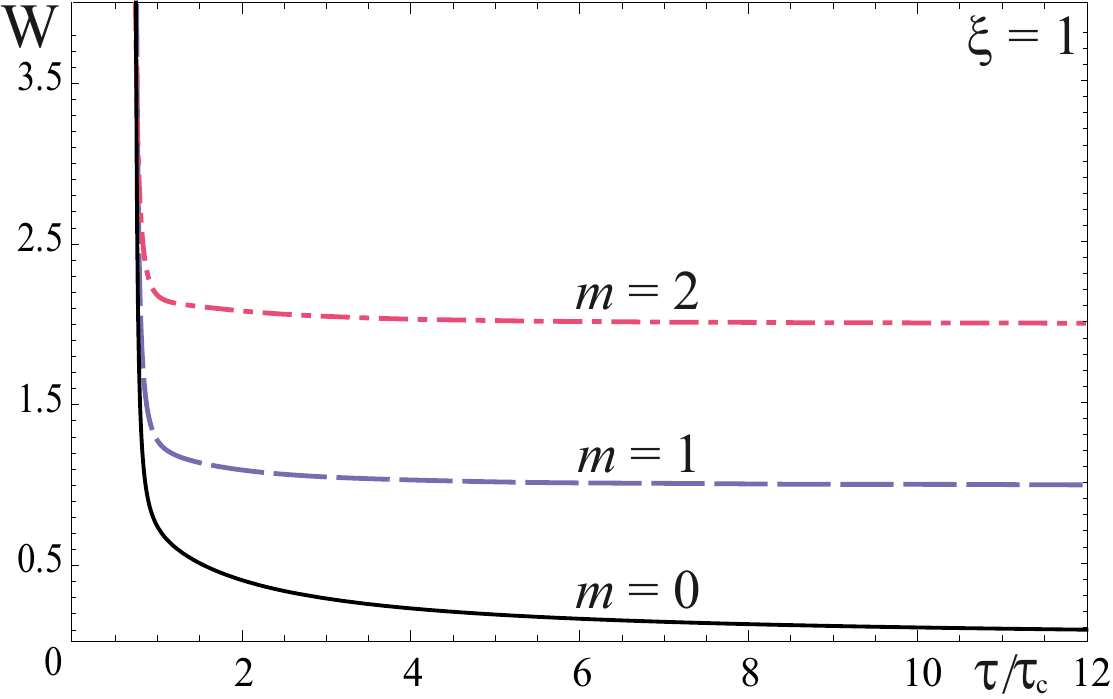}
\caption{The solution  to (\ref{eq2014}) with (\ref{eq2108}), (\ref{eq2208}),
and (\ref{eq2010}), for $k =\sqrt{3/4}$, various masses,  and 
$\xi = 0$ (top left), $\xi=1/6$ (top right), $\xi=1/3$ (bottom left), and $\xi=1$ (bottom right).
$W$ and $m$ are expressed in units of $\tau_{\rm c}^{-1}$.}
\label{fig1}
\end{center}
\end{figure*}

\subsection{Results}
\label{results}
Instead of solving equation (\ref{eq2007}), for our purpose, it is convenient to solve
equation (\ref{eq2014}) for the WKB function $W(\tau)$.
With the help of the substitution
\begin{equation}
Y(\tau)=W^{-1/2}(\tau),
\end{equation}
from (\ref{eq2014}) we obtain the differential equation
\begin{equation}
Y''+\Omega^2 Y(\tau)-Y^{-3}=0,
\label{eq4001}
\end{equation}
where the function $\Omega(\tau)$ is defined in (\ref{eq2010}) with (\ref{eq2011}).
Before proceeding to solve (\ref{eq4001}) for the analog cosmological model defined by
the metric (\ref{eq243}) with (\ref{eq2108}) and (\ref{eq2208}), it is useful to 
study three important examples
which may be solved analytically.

Consider first a conformally invariant field theory, i.e., for $m=0$ and $\xi=1/6$,
in which case there should be no particle creation, as has been argued on general grounds
\cite{parker4}.
Indeed, 
in this case, as may easily be verified, the function
\begin{equation}
Y_{\rm conf}=\omega^{-1/2}=\left(k\frac{b}{a}\right)^{-1/2}
\end{equation}
is a solution to (\ref{eq4001}) for arbitrary  $a(\tau)$ and $b(\tau)$. This  yields $W_{\rm c}(\tau)=\omega(\tau)$
and $\omega^{(n)}=0$ at all adiabatic orders $n > 0$.
Then, with the choice $\tilde{W}=W^{(n)}=\omega$,
 by virtue of (\ref{eq2033}) we find $|\beta_k|=0$ and hence no particle creation as expected.
In contrast, on account of (\ref{eq2032}) the choice $\tilde{W}=\Omega$  would  generally yield $|\beta_k|\neq 0$ and
hence an unphysical prediction of particle creation for a conformally invariant field.

Second, consider the asymptotic future. In the limit $\tau\rightarrow \infty$ our system approaches 
the zero temperature
regime and  
the spacetime described by the analog metric (\ref{eq243}) approaches the Milne universe
with $a=\tau$ and  $b=1$. 
It is therefore instructive to compare the analog cosmological particle creation with that of the Milne universe.
In particular, in the Milne universe there should be no  creation of massless particles  
since the scalar field 
satisfies the conformally invariant wave equation.
Indeed, for $m=0$ the asymptotic solution to (\ref{eq4001}) 
\begin{equation}
Y_{\infty}=\left(\frac{\tau}{k}\right)^{1/2}, \quad W_{\infty}=\omega=\frac{k}{\tau},
\end{equation}
 gives $|\beta_k|^2=0$,  and hence there is no creation of massless particles as in the previous case.
 
 Third,
it is worth analyzing the solution to (\ref{eq4001}) 
in the critical regime $t\equiv\tau- \tau_{\rm c}\rightarrow 0$.
In that regime
 $\Omega^2 \simeq \alpha t^{-2}$
where
\begin{equation}
 \alpha =6\xi \left(\beta+\beta^2-\nu/4-\nu^2/16\right)
 -\beta-\beta^2 .
 \label{eq4104}
\end{equation}
Equation (\ref{eq4001}) then simplifies to
\begin{equation}
 \frac{d^2Y}{dt^2}+\alpha  t^{-2}Y-Y^{-3}=0
 \label{eq4103}
\end{equation}
and may be solved analytically in the limit $t\rightarrow 0$.
The behavior of the solution in that limit depends crucially on 
the value of the nonminimal coupling constant
$\xi$. We find three distinct solutions depending on $\alpha$
\begin{equation}
Y=\left\{ \begin{array}{ll}
(\alpha-1/4)^{-1/4} t^{1/2}, & \mbox{ for $\alpha>1/4$},\\
 Y_1 t^{1/2-\sqrt{1-4\alpha}/2},& \mbox{for $0<\alpha<1/4$},\\
Y_2 t^{-1/2-\sqrt{1-4\alpha}/2}, & \mbox{for $\alpha<0$} , \end{array} \right.
\label{eq4105}
\end{equation}
where $Y_1$ and $Y_2$ are arbitrary real constants.
Clearly, $Y\rightarrow 0$ for $\alpha>0$ and  $Y\rightarrow \infty$ for $\alpha<0$.
From the definition (\ref{eq4104}) it follows
$\alpha \geq 0$ ($\leq 0$) if $\xi \geq \xi_{\rm c}$  ($\leq \xi_{\rm c}$) where 
\begin{equation}
 \xi_{\rm c} = \frac16 \frac{\beta+\beta^2}{\beta+\beta^2-\nu/4-\nu^2/16} 
 \label{eq4106}
\end{equation}
is the critical nonminimal coupling which takes the value
$\xi_{\rm c}\simeq 0.28322$ for the O(4) critical exponents.
Hence, the WKB function $W(\tau)$ goes to zero at the critical point if $\xi<\xi_c$ and diverges
if $\xi >\xi_{\rm c}$. Note that the critical coupling  $\xi_{\rm c}>\xi_{\rm conf}=1/6 $ 
contrary to what one would expect since the original sigma model becomes conformally 
invariant at the critical point.
Curiously,  $\xi_c$ would be equal to 
$\xi_{\rm conf}=1/6$ if the critical exponent $\nu$ were equal to zero, in which case
the pion velocity, as given by (\ref{eq252}), would not necessarily vanish at the critical point.

Since there is no creation of pions in the limit $\tau\rightarrow \infty$,
it is natural to choose as the initial state  the  vacuum state  vector $|\;\rangle_0$
at some large $\tau_0$ and evolve equation (\ref{eq2014}) backward in time 
starting at $\tau_0$. The vacuum state satisfies
\begin{equation}
 a_J|\;\rangle_0 = 0,
\quad
 a_J^\dag |\;\rangle_0=|J\rangle 
 \label{eq4002}
\end{equation}
and, according to (\ref{eq3001}), the one-particle state $|J\rangle$ is represented by 
\begin{equation}
 \langle \mbox{\boldmath{$x$}}|J\rangle= 
 \left(\frac{1}{2\tau_0^3}\right)^{1/2}\chi_k^{(0)}(\tau_0)\Phi_J(\mbox{\boldmath{$x$}}) ,
 \end{equation}
 where the function $\chi_k^{(0)}(\tau)$ is defined in (\ref{eq2020}) with $W^{(0)}=\omega$.
Then, the initial conditions consistent with  (\ref{eq2034}) are
\begin{equation}
Y(\tau_0)=\omega(\tau_0)^{-1/2}, \quad 
Y'(\tau_0)=-\frac12 \omega'(\tau_0)\omega(\tau_0)^{-3/2} .
\label{eq4003}
\end{equation}

\begin{figure*}[t]
\begin{center}
\includegraphics[width=0.45\textwidth ]{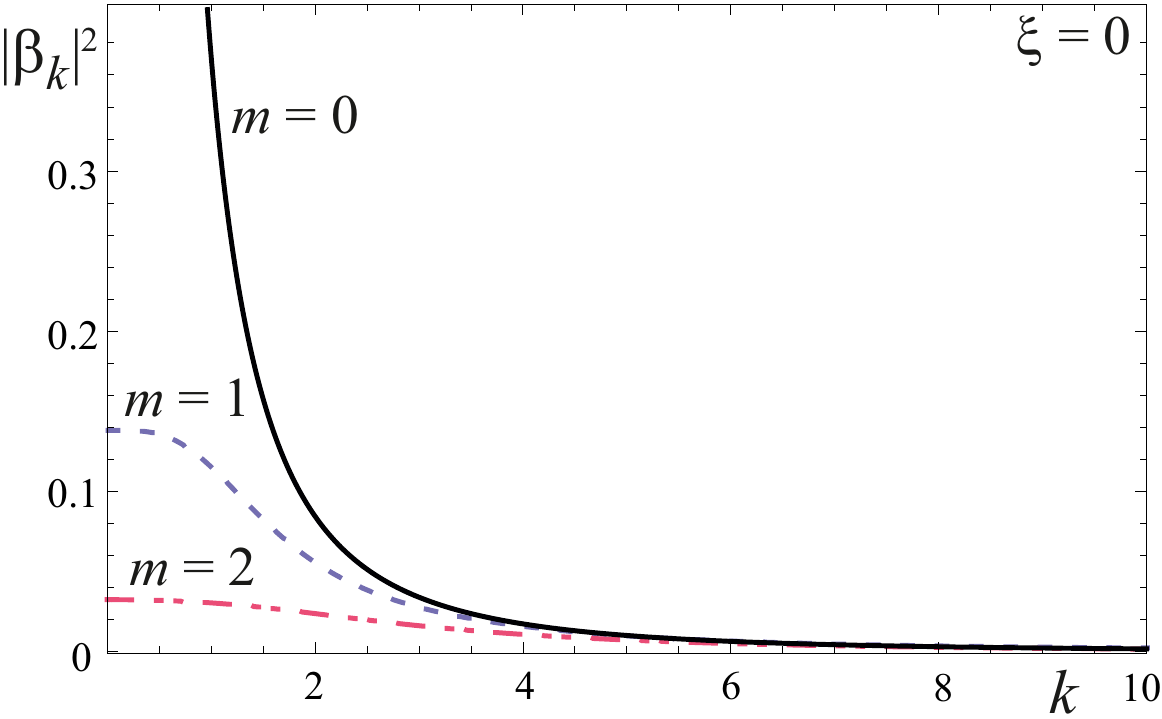}
\hspace{0.02\textwidth}
\includegraphics[width=0.45\textwidth]{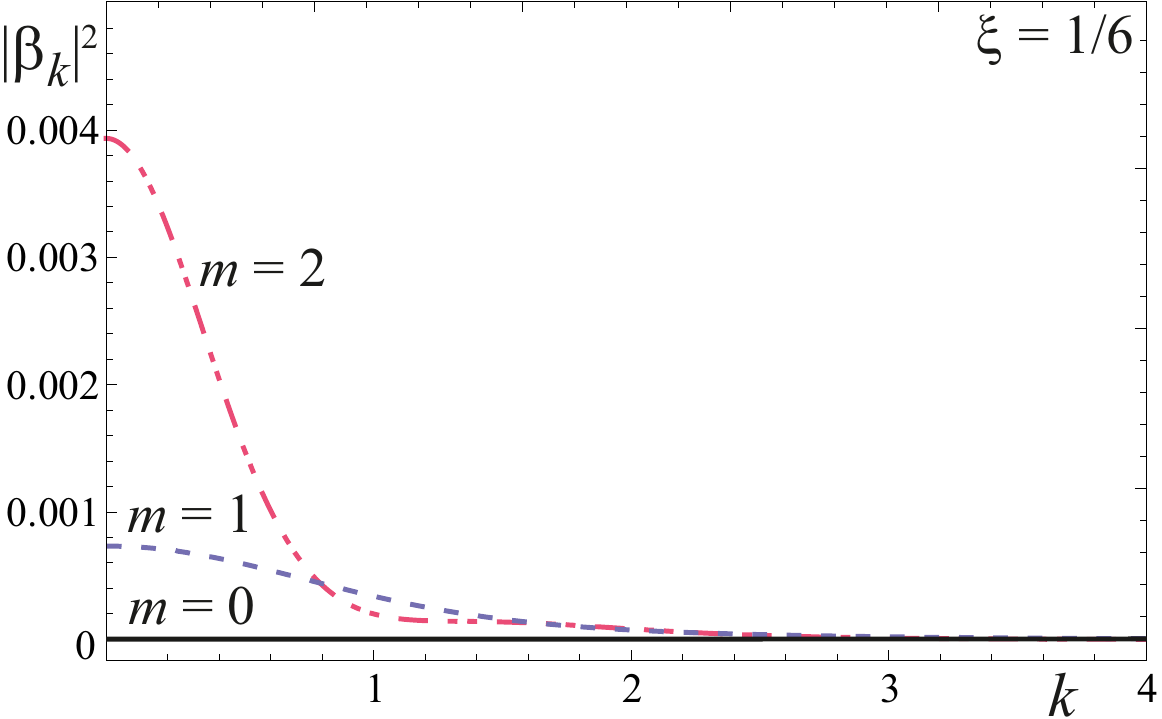}
\\
\vspace{0.02\textwidth}
\includegraphics[width=0.45\textwidth]{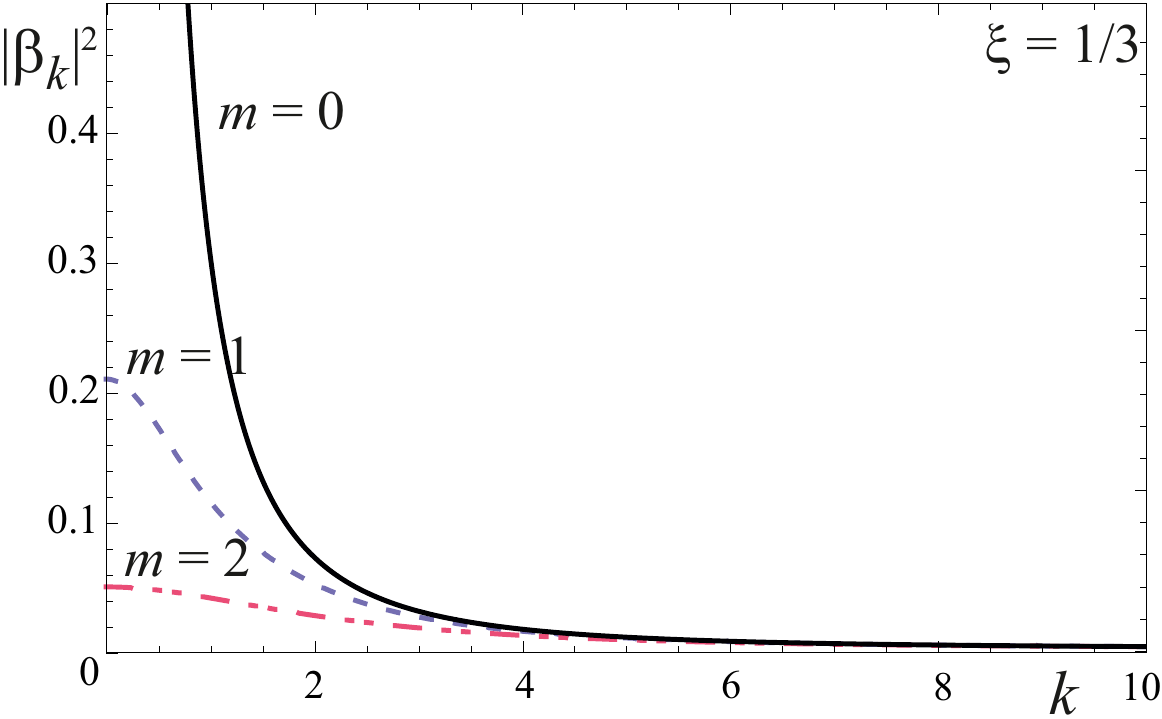}
\hspace{0.02\textwidth}
\includegraphics[width=0.45\textwidth]{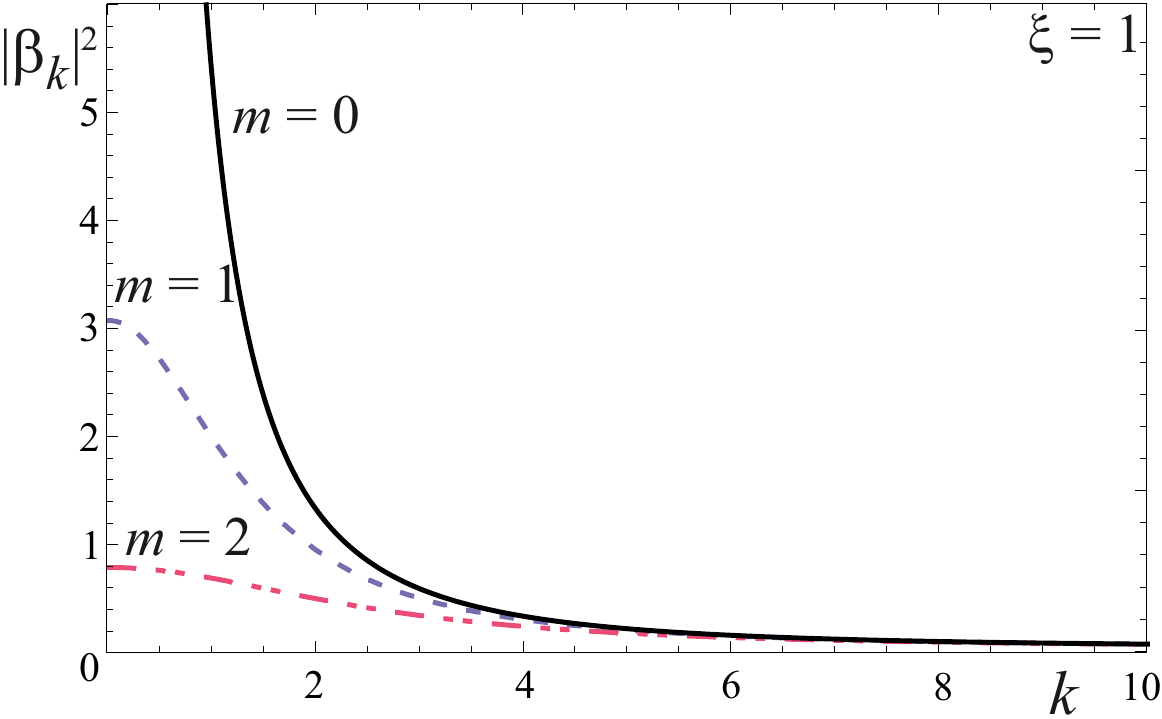}
\caption{The square of the Bogoliubov coefficient $|\beta_k|^2$ as a function of $k$
for fixed $\tau/\tau_{\rm c} =1.12$, various masses, and  $\xi = 0$ (top left), $\xi=1/6$ (top right), 
$\xi=1/3$ (bottom left), and $\xi=1$ (bottom right).}
\label{fig2}
\end{center}
\end{figure*}

\begin{figure}[t]
\begin{center}
\includegraphics[width=0.45\textwidth,trim= 0 0cm 0 0cm]{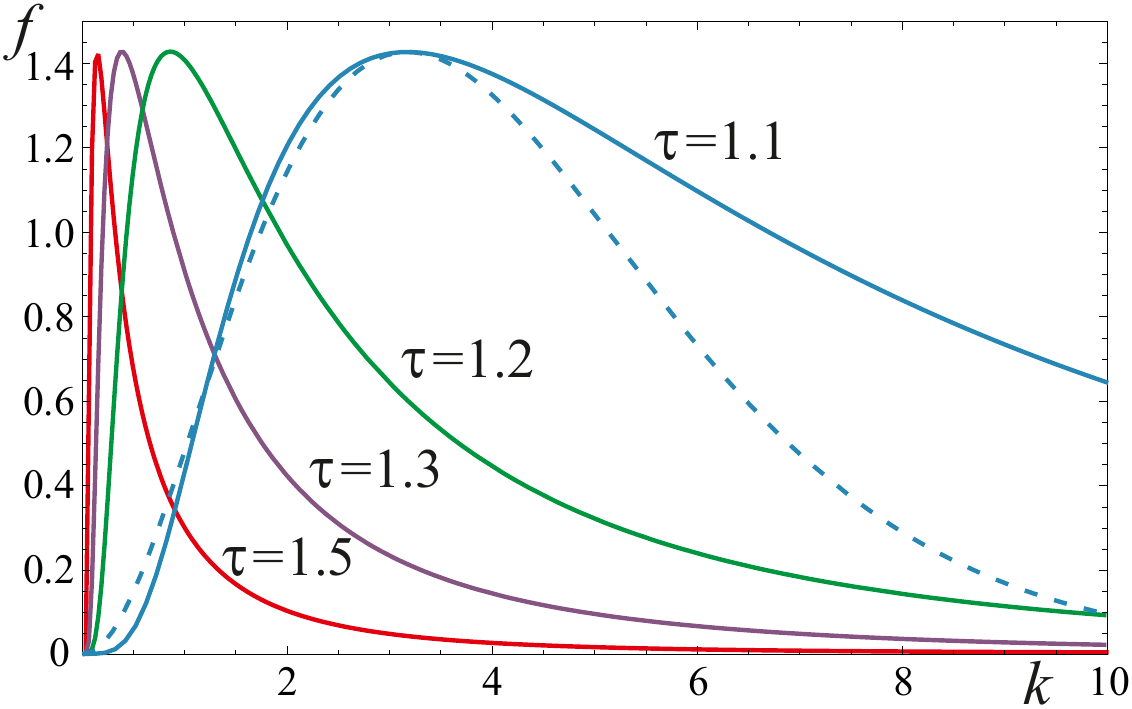}
\caption{The spectral function $f(k)$ for  $\xi = 0$, 
and various $\tau$ expressed in units of $\tau_{\rm c}$.
The dashed line represents the Planck spectral function $f_{\rm P}(k)$.}
 \label{fig3}
 \end{center}
\end{figure}

\begin{figure}[t]
\begin{center}
\includegraphics[width=0.45\textwidth,trim= 0 0cm 0 0cm]{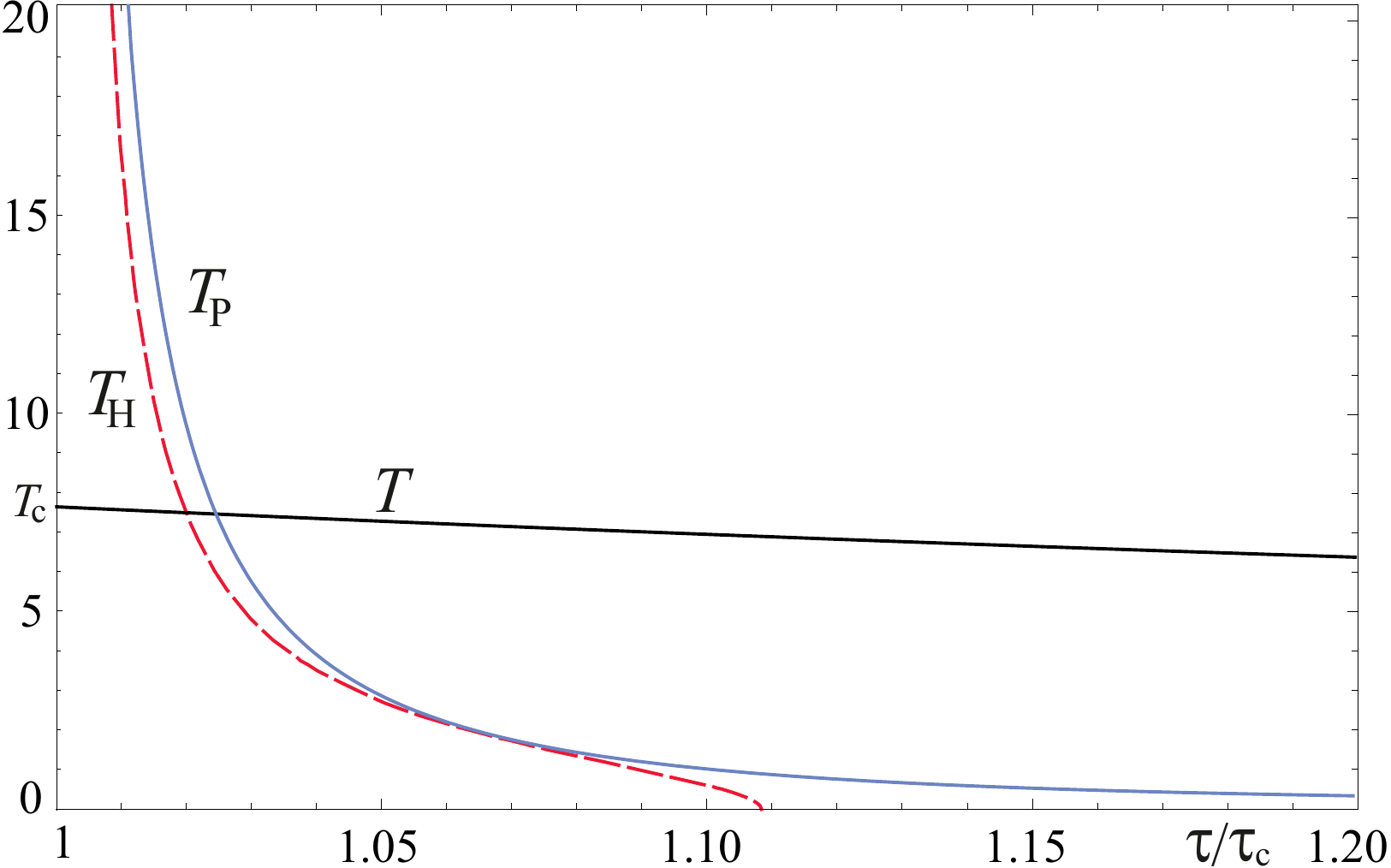}
\caption{The quasi-Planckian temperature $T_{\rm P}$ as a function of $\tau$ for  $\xi = 0$ compared with the Hawking temperature
$T_{\rm H}$ and the temperature of the fluid $T=T_{\rm c}\tau_{\rm c}/\tau$. The temperatures are plotted in  units of $\tau_{\rm c}^{-1}$.
The label $T_{\rm c}$ at 7.625 on the vertical axis marks the critical temperature.
}
 \label{fig4}
\end{center}
\end{figure}

The results of the numerical calculations are presented in Figs.\ \ref{fig1} to \ref{fig4}.
According to our conventions the comoving momentum $k$ is dimensionless,
 the time is expressed in units of $\tau_{\rm c}$,
and the mass and temperature are in units of $\tau_{\rm c}^{-1}$.
The proper time scale has been estimated in Sec.\ \ref{bjorken} from 
the phenomenology of high energy collisions  yielding
a typical value
$\tau_{\rm c}\approx 8.2 \; {\rm fm} =41.6\; {\rm GeV}^{-1}$, so
the mass scale is typically $\tau_{\rm c}^{-1} \approx 24$ MeV.

Numerical solutions to (\ref{eq2014}) with (\ref{eq2108}), (\ref{eq2208}),
and (\ref{eq2010}), $W(\tau)$, are presented in Fig.\ \ref{fig1} as functions of $\tau$
for the masses $m=0,1,2$ (in units $\tau_{\rm c}^{-1}$) and various couplings $\xi$.
We use the initial conditions (\ref{eq4003}) at $\tau/\tau_{\rm c}=12$.
Note a drastically  different behavior  
of the  $\xi=1/3$ and $\xi=1$  solutions (two bottom left panels)
with respect to the other solutions. 
The reason for that 
is the the sign change of $\Omega^2$ when $\xi$ exceeds the critical value
$\xi_c=0.28322$, as predicted by the asymptotic solution (\ref{eq4105}) 
in the vicinity of the critical point.
Using these numerical solutions we calculate the square of the Bogoliubov coefficients $|\beta_k|^2$
as given by (\ref{eq2033}). In Fig.\ \ref{fig2}, we  present $|\beta_k|^2$ 
as a function of $k$ for a fixed $\tau/\tau_{\rm c} = 1.12 $ and various couplings $\xi$.

We will now  
use the functional form of the Planck (Bose--Einstein) particle distribution function
to express 
\begin{equation}
 |\beta_k|^2 =\frac{1}{e^{E/T_{\rm P}}-1},
 \label{eq4004}
 \end{equation}
 where $E$ is the particle energy defined by (\ref{eq2104}) The ``temperature'' $T_{\rm P}$ 
depends on  $\tau$ and is generally a function of $k$.
We will call the function (\ref{eq4004}) the  {\em quasi-Planckian distribution} and 
$T_{\rm P}$ the {\em quasi-Planckian temperature}.
If $T_{\rm P}$ for a fixed $\tau$ were independent of $k$, 
the distribution function  $|\beta_k|^2$ as a function of $k$   would  
have the exact Planck form (\ref{eq4004}), and
the spectrum of created particles would be thermal.
In this case, we would be in an exact adiabatic regime \cite{barcelo1}.
If the quasi-Planckian temperature weakly depends on $k$,
 the spectrum will be nearly thermal. 
At the moment, we will assume that $T_{\rm P}$ vary weakly with $k$ and check the 
thermalization and adiabaticity
{\em a posteriori}. 

We now assume that the created pions are massless in which case $E=k/a$.
To extract $T_{\rm P}$, it is convenient to use the energy density distribution function
 \cite{landau} 
\begin{equation}
 f(k)=\frac{2\pi^2}{a^3T_{\rm P}^3} \frac{d\rho_k}{dk},
 \label{eq4005}
\end{equation}
which we have normalized so that it depends only on the dimensionless variable $x\equiv k/(aT_{\rm P})$.  
For an ideal massless boson gas
\begin{equation}
 f(k)=\frac{x^3}{e^x-1},
 \label{eq4006}
\end{equation}
with a maximum at
\begin{equation}
 x_0=2.822 .
 \label{eq4007}
\end{equation}
Using (\ref{eq4004}) we can express $f(k)$ as
\begin{equation}
 f(k)=|\beta_k|^2 \ln( 1+1/|\beta_k|^2)^3 
 \label{eq4008}
\end{equation}
and plot the right-hand side as a function of $k$ 
for various $\tau$ (Fig.\ \ref{fig3}).
Then, from the position of the maximum $k_{\rm max}(\tau)$ we obtain 
\begin{equation}
 T_{\rm P}(\tau)=\frac{k_{\rm max}(\tau)}{x_0 a(\tau)} .
 \label{eq3002}
\end{equation}

For the purpose of comparison, in Fig.\ \ref{fig3}, we  also plot
the exact Planck  
spectral function $f_{\rm P}(k)$   
 given by (\ref{eq4006}) with a $k$-independent temperature  $T_{\rm P}$.
The temperature $T_{\rm P}$ is determined from (\ref{eq3002}) so that $f_{\rm P}(k)$ coincides with our spectral function 
$f(k)$ at $k=k_{\rm max}$
 corresponding to
$\tau=1.1$. 
A comparison between $f(k)$ 
and  $f_{\rm P}(k)$  (Fig.~\ref{fig3})
shows that our quasi-Planckian spectrum is close to the Planckian
for the wave numbers near $k_{\rm max}$  and left from $k_{\rm max}$.
Clearly, the departure of $f(k)$ from  $f_{\rm P}(k)$ is 
significant for large $k$ because, according to (\ref{eq2040}),
$f(k)$ falls off asymptotically as $k^{-4}(\ln{k})^3 $  in contrast to the 
exponential decay of $f_{\rm P}(k)$.
Hence,  the calculated quasi-Planckian temperature $T_{\rm P}$ may be regarded as reliable
for almost all particles with wave numbers $k \lesssim k_{\rm max}$ and some particles with $k$ above
$k_{\rm max}$.
The number of particles thermalized at $T_{\rm P}$ may be estimated by performing the integral in
(\ref{eq2030}) with the exact Planck distribution function
 given by (\ref{eq4004}) with a $k$-independent temperature  $T_{\rm P}$.
 The excess of unthermalized particles above $k_{\rm max}$ may be estimated
 evaluating the integral from $k_{\rm max}$ to $\infty$
 using the same Planck distribution subtracted from the asymptotic expression (\ref{eq2040}).  
We find that the  proportion of such particles to the total number of thermalized ones
amounts to less than $20\%$ at $\tau=1.1$ and less than $25\%$ at $\tau=1.05$.

To check how close we are to the adiabatic regime, for each $\tau$
we compare  the comoving wave number $k$ with
the Hubble scale $|H|$. As usual, we distinguish three regimes: 
the adiabatic regime in which $k \gg  |H|$, the sudden regime in which $k \ll  |H|$,
and the intermediate regime in which $k \approx |H|$ \cite{barcelo1}.
For large $\tau$ we have $H\simeq 1/\tau$ and we are clearly in the adiabatic regime for almost all wave numbers
 since the criteria $k\gg 1/\tau$ is easily met.
However, we expect a departure from adiabaticity in the 
limit $\tau \rightarrow \tau_{\rm c}$ since in this limit
 $H$ diverges according to 
 (\ref{eq2308}). For example, 
 for $\tau=1.1$  we find  $|H|\simeq 3$, so the wave numbers satisfying $k\lesssim 3$ are not in the adiabatic regime.
 In the region of $\tau$
 between 1.5 and 1.1 depicted in Fig.~\ref{fig3}, using (\ref{eq2308}) we find $0.3 \lesssim |H| \lesssim 3$ corresponding to
 $0.4 \lesssim k_{\rm max} \lesssim 3$, so our estimated $k_{\rm max}$ are of the order of $|H|$ 
 and fall into the intermediate regime.

In Fig.\ \ref{fig4}, we plot the quasi-Planckian temperature as a function of $\tau$
for vanishing nonminimal coupling constant $\xi$ together with the Hawking temperature $T_{\rm H}$
of thermal pions emitted at the apparent horizon. For comparison we plot in the same figure
the background temperature of the fluid vs $\tau$
as given by equation (\ref{eq007}) with $T_{\rm c}=$ 7.625 $\tau_{\rm c}^{-1}$.

The functional dependence of $T_{\rm H}$ is calculated 
following the prescription of our previous papers \cite{tolic,tolic2}.
As shown in Ref.\  \cite{tolic2}, the condition that a two-dimensional surface $H$
is the apparent horizon for a spherically symmetric spacetime
 may be expressed as
\begin{equation}
G^{\mu\nu}n_\mu n_\nu|_H=0,
\label{a1}
\end{equation}
where  $n_\mu$ is a vector field  normal to the surfaces of spherical symmetry. 
For the metric (\ref{eq243}) the vector $n_\mu$ is given by
\begin{equation}
n_\mu=\partial_\mu (a \sinh y).
\label{a2}
\end{equation} 
Using this  and (\ref{a1}) we obtain the condition for the analog apparent horizon 
in the form
\begin{equation}
\frac{\dot{a}}{b}\pm \frac{1}{\tanh y}=0.
\label{a3}
\end{equation}
Provided this condition is met, the surface gravity $\kappa$ at the horizon 
may be calculated using the Kodama--Hayward prescription
\cite{hayward2} which we have adapted to analog gravity \cite{tolic}.
This prescription involves theso-called Kodama vector $K^\mu$ \cite{kodama} which generalizes 
the concept of the time translation Killing vector to nonstationary spacetimes.
The analog surface gravity $\kappa$ is defined by 
\begin{equation}
\kappa =\frac{1}{2} \frac{1}{\sqrt{-h}} \partial_\alpha ( \sqrt{-h}h^{\alpha\beta}K n_\beta),
\label{eq228}
\end{equation}
where the quantities on the right-hand side should be evaluated at the apparent horizon.
 The tensor $h^{\alpha\beta}$ is the inverse of the metric $h_{\alpha\beta}$
 of the two-dimensional space normal to the surface of spherical symmetry
 and $h=\det h_{\alpha\beta}$.
 The definition (\ref{eq228})  differs from the original 
expression for the dynamical surface gravity \cite{hayward2}
by   a normalization factor $K$  which we have introduced in order to meet the requirement that $K^\mu$
should coincide with the time translation Killing
vector $\xi^\mu$  for a stationary geometry.
The details of the calculation and the final expression for $\kappa$
may be found in Ref.\ \cite{tolic}.
The corresponding temperature $T_{\rm H}=\kappa/(2\pi)$ represents 
the analog Hawking temperature of thermal pions emitted at the apparent horizon.

As shown in Ref.\ \cite{tolic}, the Hawking temperature diverges near the critical point as 
\begin{equation}
T_{\rm H} \propto (\tau-\tau_{\rm c})^{-1} .
 \label{eq4009}
\end{equation}
The temperature $T_{\rm P}$ seems to diverge at the critical point in a similar way 
and vanishes in the limit 
$\tau \rightarrow \infty$ corresponding to the zero background temperature
of the hadronic fluid. In contrast,
the analog Hawking temperature vanishes at $\tau=\tau_{\rm max}=$ 1.1002 $\tau_{\rm c}$ at which the
analog trapping horizon ceases to exist \cite{tolic2}.

\section{Conclusions}
\label{conclusion}

We have investigated the cosmological creation of pions 
in an expanding hadronic fluid 
 in the regime near the critical point of the chiral phase transition.
 In our approach we have disregarded a possible particle production caused by
 the self-interaction potential of the scalar field.
 Besides, we have assumed that the created pions are of zero or 
very light mass and we have neglected the creation of much heavier sigma mesons.
 The production rate has been calculated using the adiabatic regularization prescription
 according to which the Bogoliubov coefficients are expressed in terms of  
the WKB function $W(\tau)$ and its first-order adiabatic approximation $W^{(1)}(\tau)=\omega(\tau)$. 
The function $W(\tau)$ has been computed by solving equation (\ref{eq2014}) numerically.
We have analyzed more closely the solution in the limit when $\tau$ approaches the critical value.
It turns out that the behavior of the solution 
in that limit
depends crucially on 
the value of the nonminimal coupling constant
$\xi$. 
We have shown that there exists a certain critical value $\xi_{\rm c}$ larger than 
the conformal value $\xi=1/6$
such that $W(\tau)$ goes to zero
at the critical point for $\xi<\xi_{\rm c}$  and diverges for $\xi>\xi_{\rm c}$.

We have calculated the cosmological production rate as a function of the proper time
for various masses and various nonminimal coupling constants $\xi$.
The production rate of massless pions shows a strong dependence on   $\xi$
and vanishes for  $\xi=1/6$ as it should.
By fitting the production rate  to  the 
Planck blackbody radiation spectrum we have extracted 
the temperature of the produced pion gas.
We use the time dependence of the thus obtained quasi-Planckian temperature to
compare the analog Hawking effect with the analog cosmological particle creation.
As we have already mentioned, these two effects, 
although being of similar quantum origin, 
are quite distinct physical phenomena that appear under different physical conditions.
Compared with the analog Hawking radiation of pions at the trapping horizon,
the spectrum of the cosmological radiation shows a similar behavior near the critical point. 
The temperature of the cosmologically created pions $T_{\rm P}$  diverges at the critical point 
roughly in the same way as the analog Hawking temperature $T_{\rm H}$. 
However, as the  proper time increases, the quasi-Planckian temperature vanishes asymptotically whereas 
the analog Hawking temperature  vanishes at 
a finite proper time of the order 
1.1 $\tau_{\rm c}$ when the analog trapping horizon disappears.  

Our results could not be  easily confronted with observations.
First of all, we are dealing with exact spherical symmetry, whereas
in most high energy collisions
the symmetry is axial involving
a transverse expansion
superimposed on a longitudinal boost invariant expansion.
Second,
the cosmologically created and Hawking radiated pions could not be 
easily distinguished from the background pions produced
directly from the quark-gluon plasma (QGP).
Nevertheless, we can draw a qualitative postcollision picture
as follows.

 The high temperature (QGP) produced in the collision 
 expands and cools down until the temperature
is as low as the deconfinement temperature of the order of $T_{\rm dec}\simeq T_{\rm c}= 183$.
Then, a hadronic fluid mainly consisting of pions forms and expands further according to
the Bjorken model with the proper time related to the background fluid temperature  through the relation
(\ref{eq007}), where  $\tau_{\rm c}^{-1}\simeq 24$ MeV.
Immediately below  $T_{\rm c}= 183$, the cosmological creation and the Hawking radiation
take place. Initially, both the Hawking and quasi-Planckian temperatures  exceed the background fluid temperature $T$
by a factor of $2$ or more. As a consequence, a considerable fraction of the pion gas
will  be briefly ``reheated'' but, according to Fig.\ \ref{fig1},
will quickly cool down  during the subsequent expansion.
Whereas the Hawking radiation stops at $\tau=\tau_{\max}$ 
when the temperature of the fluid is of the order 0.9 $T_{\rm c}$,
the cosmological creation continues up to the thermal freez-out.
The thermal (or kinetic) freez\-out takes place  soon after the so-called chemical freez-out
which is very close or equal to the QCD deconfinement transition
\cite{braun}. 
The kinetic freez-out temperature depends  on the collision energy
\cite{star} and is roughly between 0.7 and  0.9 $T_{\rm c}$,
which corresponds to
the proper time interval
$(1.1,1.4)$.
From Fig.\ \ref{fig4} it is evident that
the influence of the cosmological production and the Hawking radiation 
will be more pronounced if the thermal freez-out is closer to the critical temperature.

\subsection*{Acknowledgments}
We  gratefully acknowledge enlightening discussions with  S.~Liberati and M.~Visser  
which motivated us to consider particle creation in the hadronic analog universe.
This work has been supported by the Croatian Science
Foundation under the project (IP-2014-09-9582) and
supported in part by the ICTP-SEENET-MTP Grant
No. PRJ-09 “Strings and Cosmology” in the frame of the
SEENET-MTP Network.

\end{document}